# From Educational Analytics to AI Governance: Transferable Lessons from Complex Systems Interventions

Hugo Roger Paz
*Faculty of Exact Sciences and Technology*
*National University of Tucumán (UNT), Argentina*
ORCID: 0000-0003-1237-7983
Contact: hpaz@herrera.unt.edu.ar

## ABSTRACT

Both student retention in higher education and artificial intelligence governance face a common structural challenge: the application of linear regulatory frameworks to complex adaptive systems. Risk-based approaches dominate both domains, yet systematically fail because they assume stable causal pathways, predictable actor responses, and controllable system boundaries. This paper extracts transferable methodological principles from CAPIRE (*Curriculum, Archetypes, Policies, Interventions & Research Environment*), an empirically validated framework for educational analytics that treats student dropout as an emergent property of curricular structures, institutional rules, and macroeconomic shocks. Drawing on longitudinal data from engineering programmes and causal inference methods, CAPIRE demonstrates that well-intentioned interventions routinely generate unintended consequences when system complexity is ignored. We argue that five core principles developed within CAPIRE—temporal observation discipline, structural mapping over categorical classification, archetype-based heterogeneity analysis, causal mechanism identification, and simulation-based policy design—transfer directly to the challenge of governing AI systems. The isomorphism is not merely analogical: both domains exhibit non-linearity, emergence, feedback loops, strategic adaptation, and path dependence. We propose Complex Systems AI Governance (CSAIG) as an integrated framework that operationalises these principles for regulatory design, shifting the central question from ‘how risky is this AI system?’ to ‘how does this intervention reshape system dynamics?’ The contribution is twofold: demonstrating that empirical lessons from one complex systems domain can accelerate governance design in another, and offering a concrete methodological architecture for complexity-aware AI regulation.

**Keywords:** complex systems, AI governance, learning analytics, causal inference, agent-based modelling, risk-based regulation, emergent harm, policy simulation

# 1. INTRODUCTION

## 1.1 The shared problem

Two domains that rarely appear in the same sentence—student retention in higher education and the governance of artificial intelligence—share a pathology so fundamental that it typically goes unnoticed: both are governed by frameworks designed for linear systems yet operate as complex adaptive systems where linearity assumptions systematically fail.

In higher education, decades of retention research have produced increasingly sophisticated models of student dropout (Tinto, 1975; Bean, 1980; Astin, 1984). Yet institutional interventions based on these models—early warning systems, academic support programmes, financial aid schemes—routinely underperform expectations. The dominant approach treats dropout as an individual-level phenomenon driven by identifiable risk factors: socioeconomic background, prior academic performance, engagement indicators. Interventions target these factors directly, assuming that addressing individual deficits will produce proportional reductions in attrition. What this framing obscures is that dropout is not merely an attribute of students but an emergent property of the interaction between students, curricular structures, institutional rules, and external shocks (Paz, 2025a). A rigid curriculum with long prerequisite chains, strict regularity requirements, and limited flexibility creates structural conditions under which even well-prepared students accumulate delays that eventually become insurmountable. The curriculum does not cause dropout in any simple sense; rather, it amplifies vulnerability and transforms minor setbacks into cascading failures.

In AI governance, a parallel pattern has emerged. The dominant paradigm—risk-based regulation—promises proportional controls matched to anticipated harms (OECD, 2019; NIST, 2023). Systems are classified into risk categories, obligations are calibrated accordingly, and compliance is monitored through documentation, audits, and procedural safeguards. The European Union’s Artificial Intelligence Act exemplifies this approach, distinguishing between unacceptable, high, and limited-risk applications and imposing graduated requirements (European Parliament and Council, 2024). Yet this framework rests on assumptions that mirror those of traditional retention research: risks can be enumerated ex ante, causal pathways are sufficiently stable to support prediction, and regulatory interventions act as external corrections to an otherwise tractable system.

Both sets of assumptions fail for the same structural reason. Complex adaptive systems do not respond to interventions through simple cause-effect chains. They adapt, reorganise, and generate emergent behaviours that evade linear models of control (Holland, 1992; Sterman, 2000; Meadows, 2008). In education, students optimise for regulatory categories rather than learning outcomes, curricula create bottlenecks that concentrate failure at specific points, and macroeconomic shocks interact with institutional rigidities in ways that cannot be anticipated from individual-level predictors alone. In AI governance, firms restructure decision pipelines to avoid regulatory thresholds, risks migrate from visible to opaque actors, and compliance becomes a performance that satisfies formal requirements while preserving underlying dynamics unchanged (Paz, 2025b).

The shared problem, then, is not insufficient data, inadequate models, or weak enforcement. It is a fundamental misalignment between the ontology of the system being governed and the logic of the governance framework applied to it. Linear frameworks assume that systems are decomposable, that interventions act locally, and that effects scale proportionally with causes. Complex adaptive systems violate all three assumptions.

### 1.2 The opportunity for transfer

If the structural problem is shared, methodological solutions developed in one domain may transfer to the other. This paper argues that educational systems offer a particularly valuable source of transferable lessons for AI governance, for three reasons.

First, educational systems function as natural laboratories for complex systems intervention. Universities generate rich longitudinal data on individual trajectories, institutional structures, policy changes, and external shocks. The effects of interventions unfold over observable timescales—semesters and years rather than decades—permitting empirical evaluation that would be prohibitively slow or politically costly in other domains. Researchers can trace how students respond to curricular changes, how institutions adapt to regulatory pressures, and how macro-level shocks propagate through organisational structures. This empirical density is largely absent from AI governance, where regulatory frameworks are too new, data too fragmented, and timescales too compressed to support comparable analysis.

Second, the stakes in educational analytics permit methodological experimentation that AI governance cannot yet afford. Testing a new retention intervention on a student cohort carries real consequences, but those consequences are bounded in ways that testing AI governance frameworks on deployed systems is not. Educational researchers can run natural experiments, compare policy regimes across institutions, and observe unintended consequences without risking the systemic harms that premature AI regulation might generate. The methodological lessons learned—what works, what fails, what generates emergent pathologies—can then inform governance design in higher-stakes domains.

Third, the structural isomorphism between educational systems and AI ecosystems is not merely metaphorical but formal. Both exhibit the defining properties of complex adaptive systems: non-linearity, emergence, feedback loops, strategic adaptation, and path dependence (Mitchell, 2009; Cilliers, 1998). Students adapt to curricular constraints just as firms adapt to regulatory constraints. Curricula create structural conditions that shape possible trajectories just as AI pipelines create structural conditions that shape possible harms. Macroeconomic shocks interact with institutional rigidities in education just as competitive pressures interact with compliance requirements in AI markets. The isomorphism justifies treating methodological principles as transferable rather than domain-specific.

### 1.3 Contribution and structure of the paper

This paper makes two contributions. The first is demonstrative: showing that empirical lessons from one complex systems domain—educational analytics—can accelerate governance design in another—artificial intelligence regulation. The second is constructive: proposing Complex Systems AI Governance (CSAIG) as an integrated framework that operationalises complexity-aware principles for regulatory design.

The vehicle for both contributions is CAPIRE (*Curriculum, Archetypes, Policies, Interventions & Research Environment*), a framework developed and empirically validated for analysing student trajectories in engineering education (Paz, 2025a). CAPIRE treats dropout not as an individual failure but as an emergent property of curricular structure, institutional rules, and external shocks. It integrates graph-theoretic analysis of curricula, unsupervised identification of trajectory archetypes, causal inference through Double Machine Learning methods, and agent-based simulation for counterfactual policy analysis. Crucially, CAPIRE was developed through iterative engagement with real institutional data, revealing both what works methodologically and what unintended consequences emerge when complexity is underestimated.

From CAPIRE, we extract five transferable principles for complex systems governance: (1) temporal observation discipline, which respects the order in which information becomes available and prevents leakage between what is known ex ante and what is learned ex post; (2) structural mapping over categorical classification, which prioritises understanding system topology over assigning static risk labels; (3) archetype-based heterogeneity analysis, which recognises that different actor types respond differently to identical interventions; (4) causal mechanism identification, which distinguishes correlation from intervention-relevant causation; and (5) simulation-based policy design, which tests interventions against adaptive responses before deployment.

The paper proceeds as follows. Section 2 establishes the theoretical foundations, arguing that complexity constitutes a shared ontology that justifies cross-domain transfer. Section 3 presents CAPIRE in sufficient detail to ground the subsequent argument, focusing on empirical findings rather than technical implementation. Section 4 extracts the five transferable principles and articulates their general form. Section 5 applies these principles to AI governance, critiquing existing frameworks and proposing CSAIG as an alternative architecture. Section 6 discusses implications for policy, institutions, and research, while acknowledging limitations of the transfer. Section 7 concludes with a reflection on what it means to govern systems whose behaviour cannot be fully predicted.

Throughout, the argument rests on a single core claim: that the central question of governance must shift. In education, the question is not ‘which students are at risk?’ but ‘how do curricular structures and institutional rules generate vulnerability, and how do interventions reshape those dynamics?’ In AI governance, the question is not ‘how risky is this system?’ but ‘how does this regulatory intervention alter the behaviour of the sociotechnical system, and what new configurations might it generate?’ Both questions demand methodological frameworks adequate to complexity. CAPIRE offers one such framework, empirically tested; CSAIG proposes its extension to a domain where the stakes are higher and the data sparser, but where the structural logic remains the same.

## 2. THEORETICAL FOUNDATIONS: COMPLEXITY AS SHARED ONTOLOGY

The claim that methodological principles can transfer between educational analytics and AI governance rests on a deeper assertion: that both domains share a common ontology. This section argues that complexity—understood not as a rhetorical label but as a precise description of system properties—constitutes that shared foundation. Once complexity is taken seriously as an analytical commitment, the structural isomorphism between the two domains becomes not merely plausible but formally demonstrable.

### 2.1 Properties of complex adaptive systems

Complex adaptive systems differ from merely complicated ones not in degree but in kind (Mitchell, 2009). A complicated system—an aircraft engine, a hospital's scheduling algorithm, a tax code—may contain many interacting components and require substantial expertise to understand, yet remains fundamentally predictable given sufficient information. Its behaviour can be decomposed, its parts analysed in isolation, and its responses to intervention anticipated with reasonable accuracy. Complex adaptive systems operate under different principles entirely.

Five properties define complex adaptive systems and distinguish them from their merely complicated counterparts (Holland, 1992; Sterman, 2000; Meadows, 2008; Cilliers, 1998).

*Non-linearity.* Effects do not scale proportionally with causes. Small perturbations may trigger cascading changes across the system, while large interventions may be absorbed with minimal impact. This property undermines governance strategies based on proportional control, where regulatory intensity is calibrated to anticipated effect size. In non-linear systems, the relationship between intervention magnitude and outcome magnitude is fundamentally unstable.

*Emergence.* System-level patterns arise from interactions among components rather than from the properties of components themselves. Emergent phenomena cannot be predicted by analysing elements in isolation; they exist only at the level of the whole. Crucially, emergence implies that harms may arise without any single component being defective or any single decision being wrong. The harm is a property of configuration, not of parts.

*Feedback loops.* Actions within the system feed back to alter subsequent conditions. Positive feedback amplifies initial changes, potentially driving the system toward new equilibria or tipping points. Negative feedback dampens change, creating stability but also resistance to intervention. Both forms complicate governance: positive feedback can transform minor regulatory gaps into systemic failures, while negative feedback can neutralise reform efforts through compensatory adaptation.

*Strategic adaptation.* Agents within the system are not passive recipients of constraints but active optimisers who learn, adjust, and reorganise in response to their environment. Regulatory interventions become part of that environment, shaping incentives and triggering behavioural responses that may reinforce, undermine, or redirect the intervention's intended effects. Over time, the system co-evolves with its governance regime.

*Path dependence and irreversibility.* Historical trajectories constrain future possibilities. Once certain configurations become established—through market dominance, infrastructural lock-in, institutional routines, or accumulated human capital—they become difficult or impossible to reverse. This property implies that regulatory timing matters as much as regulatory content: interventions that might succeed early may fail once path-dependent structures have solidified.

These five properties are not independent. They interact to produce the characteristic unpredictability of complex systems. Non-linearity makes feedback effects difficult to anticipate; emergence ensures that adaptive responses generate novel configurations; path dependence means that the effects of today's interventions constrain tomorrow's options. Together, they define an ontology in which linear models of cause and effect are not merely imprecise but categorically inappropriate.

### 2.2 The educational system as a complex adaptive system

Higher education systems exhibit all five properties in forms that are empirically observable and consequential for policy.

Students function as adaptive agents who respond strategically to institutional constraints. They optimise course loads to satisfy regularity requirements, sequence enrolments to manage workload, and adjust effort allocation based on perceived stakes (Tinto, 1993). These responses are rational given local incentives but may generate aggregate outcomes—grade inflation, bottleneck congestion, strategic course avoidance—that no individual intended. The curriculum does not merely present knowledge; it structures a decision environment in which students navigate trade-offs between competing demands.

Curricula themselves constitute constraint structures that shape possible trajectories. A curriculum with long prerequisite chains, strict regularity requirements, and limited course availability creates structural conditions under which delays compound. Missing a single gateway course may push back an entire sequence; losing regularity status may trigger cascading constraints that foreclose previously available paths. These effects are emergent: they arise not from the difficulty of any single course but from the interaction between curricular topology and student behaviour under uncertainty.

Institutions respond adaptively to both student behaviour and external pressures. Departments adjust offerings based on enrolment patterns; administrators modify policies in response to retention metrics; accreditation requirements reshape programme structures. These institutional adaptations interact with student adaptations in feedback loops that may stabilise the system around suboptimal equilibria or, under stress, trigger rapid reorganisation.

External shocks—economic crises, labour market shifts, public health emergencies, industrial action—propagate through the system in ways that linear models cannot capture. A teachers' strike does not merely delay instruction; it interacts with curricular rigidity to convert temporary disruption into permanent attrition for students who cannot absorb the resulting friction (Paz, 2025a). Inflation does not merely reduce purchasing power; it reshapes time allocation between study and work, alters opportunity costs of persistence, and shifts the distribution of who can afford to remain enrolled. These macro-level forces do not act uniformly; they interact with structural vulnerabilities to produce heterogeneous effects across student populations.

Path dependence manifests in both individual trajectories and institutional structures. A student who falls behind in early semesters carries accumulated disadvantage forward; recovery becomes progressively more difficult as delays compound and deadlines approach. At the institutional level, curricula developed decades ago persist through faculty hiring patterns, laboratory infrastructure, and accreditation inertia, even when the original rationale has become obsolete.

### 2.3 The AI ecosystem as a complex adaptive system

The ecosystem surrounding artificial intelligence exhibits the same five properties, manifested in different but structurally parallel forms.

Developers, deployers, and users function as adaptive agents who respond strategically to regulatory and market constraints. Firms restructure decision pipelines to minimise regulatory exposure, distribute functionality across jurisdictions, modularise systems to avoid classification thresholds, and substitute regulated practices with functionally equivalent but formally distinct alternatives (Veale & Zuiderveen Borgesius, 2021). These responses are rational given competitive pressures but may generate aggregate outcomes—regulatory arbitrage, risk displacement, compliance theatre—that undermine governance objectives.

AI systems themselves constitute constraint structures that shape possible outcomes. A model trained on particular data, optimised for particular objectives, and deployed in particular contexts creates structural conditions under which certain harms become more or less likely. These conditions interact with deployment environments, user behaviour, and institutional contexts in ways that cannot be anticipated from model properties alone. Discriminatory outcomes, for instance, may emerge not from biased training data but from the interaction between model behaviour, selection effects in deployment, and adaptive responses by affected populations (Barocas & Selbst, 2016).

Markets and institutions respond adaptively to both technological capability and regulatory pressure. Investment flows toward applications with favourable regulatory treatment; talent migrates toward organisations with compliance capacity; industry standards emerge through competitive dynamics that may or may not align with public interest. These institutional adaptations interact with firm-level adaptations in feedback loops that shape the evolution of the technology itself.

External shocks—geopolitical tensions, computational breakthroughs, viral deployment failures, shifts in public sentiment—propagate through the ecosystem in ways that linear models cannot capture. A single high-profile failure may trigger regulatory cascades; a breakthrough capability may reshape competitive dynamics overnight; geopolitical conflict may fragment previously unified markets. These shocks interact with structural conditions to produce effects that are heterogeneous across actors and contexts.

Path dependence manifests in both technical trajectories and market structures. Architectural choices made early in development constrain subsequent possibilities; data accumulated by first movers creates barriers to entry; regulatory frameworks designed for current capabilities may be poorly suited to emerging ones. The AI ecosystem, like the educational system, carries its history forward in ways that constrain present options and future trajectories.

### 2.4 Structural isomorphism and the justification for transfer

The parallel exhibition of complex system properties in both domains is not coincidental. Both are sociotechnical systems in which heterogeneous agents operate under constraint structures, adapt strategically to incentives, and generate emergent outcomes through interaction (Jasanoff, 2004). The specific content differs—students versus firms, curricula versus models, academic regulations versus AI governance—but the structural logic is the same.

Table 1 summarises this isomorphism across the five defining properties of complex adaptive systems.

**Table 1.** Structural isomorphism between educational systems and AI ecosystems

| **Property** | **Educational systems** | **AI ecosystems** |
|---|---|---|
| **Non-linearity** | Small curricular changes trigger disproportionate retention effects; large reforms absorbed with minimal impact | Minor regulatory changes trigger major compliance restructuring; extensive rules neutralised through adaptation |
| **Emergence** | Dropout arises from student–curriculum–institution interaction, not individual deficits | Harm arises from model–deployment–context interaction, not model properties alone |
| **Feedback loops** | Retention policies reshape student behaviour; student behaviour reshapes institutional response | Regulation reshapes firm behaviour; firm behaviour reshapes effective meaning of regulation |
| **Strategic adaptation** | Students optimise for regulatory categories (regularity, credits) over learning objectives | Firms optimise for compliance categories (risk classification) over harm reduction |

| Property | Educational systems | AI ecosystems |
|---|---|---|
| **Path dependence** | Early delays compound into structural disadvantage; curricular inertia persists across decades | Early architectural choices constrain development; market concentration creates entry barriers |

This isomorphism justifies treating methodological principles as transferable. If both domains are complex adaptive systems, then frameworks developed to navigate complexity in one domain should, in principle, apply to the other. The transfer is not analogical but structural: it rests not on surface similarities but on shared deep properties that determine how systems respond to intervention.

The justification has limits, of course. The domains differ in scale, in the nature of their agents, in the observability of outcomes, and in the political economy of intervention. Section 6 addresses these limitations directly. But the core claim stands: complexity constitutes a shared ontology that makes cross-domain transfer not merely plausible but methodologically grounded.

## 3. THE CAPIRE FRAMEWORK: EMPIRICAL FOUNDATIONS

Having established complexity as the shared ontology linking educational systems and AI ecosystems, we now turn to CAPIRE: the framework that operationalises complexity-aware analysis in the educational domain. This section presents CAPIRE in sufficient detail to ground the subsequent extraction of transferable principles, focusing on empirical findings and methodological architecture rather than technical implementation.

### 3.1 Context and original problem

CAPIRE was developed to address student attrition in engineering programmes at the National University of Tucumán (UNT), Argentina. The context presents a particularly demanding test case for complexity-aware methods: a rigid curriculum with extensive prerequisite chains, strict regularity requirements that constrain enrolment options, limited course offerings that create structural bottlenecks, and an unstable macroeconomic environment characterised by recurring inflation spikes and periodic teachers' strikes.

Traditional retention interventions at UNT—tutoring programmes, early warning systems, financial aid—had produced disappointing results. The interventions were well-designed by conventional standards, targeting known risk factors with evidence-based approaches. Yet attrition rates remained stubbornly high, and the distribution of dropout across cohorts exhibited patterns that individual-level models could not explain. Students with similar entry profiles experienced dramatically different outcomes; cohorts entering in different years showed divergent trajectories despite comparable preparation; and the effects of support interventions varied unpredictably across contexts.

These anomalies suggested that the dominant framing—dropout as individual deficit—was missing something structural. CAPIRE emerged from the hypothesis that dropout is better understood as an emergent property of the interaction between students, curricular structures, institutional rules, and external shocks. If this hypothesis is correct, then effective intervention requires understanding system dynamics, not merely identifying at-risk individuals.

### 3.2 Architecture of the framework

CAPIRE is organised as an integrated ecosystem comprising three layers: a data layer that structures longitudinal information about students, institutions, and contexts; an analytics layer that extracts patterns, estimates causal effects, and identifies structural vulnerabilities; and a simulation layer that tests policy interventions against adaptive responses before deployment.

#### 3.2.1 Data layer: multi-level architecture and temporal discipline

The data layer implements a four-level architecture (N1–N4) that organises information by the temporal order in which it becomes available:

- *N1 (Pre-entry context):* Information available before a student enters the programme—secondary school background, socioeconomic indicators, geographical origin, family educational history. These variables are strictly exogenous to the university experience.
- *N2 (Entry moment):* Information generated at the point of matriculation—entry cohort, initial programme choice, admission pathway, diagnostic assessment results. These variables establish the initial conditions for the student's trajectory.
- *N3 (Academic trajectory):* Information accumulated during enrolment—course attempts and outcomes, credit accumulation, regularity status, curricular friction metrics. These variables evolve over time and reflect the interaction between student behaviour and institutional constraints.
- *N4 (External context):* Information about macro-level conditions—inflation rates, teachers' strike duration, labour market indicators, policy changes. These variables are exogenous to individual students but interact with institutional structures to shape aggregate outcomes.

The multi-level architecture enforces a principle termed *Value of Observation Time* (VOT): variables are only included in predictive or causal models at points where they would actually be observable. This prevents temporal leakage—the inadvertent use of future information to predict past outcomes—which is a pervasive source of inflated performance estimates in educational data mining (Aguiar et al., 2015). More importantly for governance, VOT ensures that insights derived from the model correspond to information that institutions could actually use for intervention.

#### 3.2.2 Analytics layer: graphs, archetypes, and causal inference

The analytics layer comprises three integrated components that progressively build understanding of system dynamics.

*Curriculum graph analysis.* The curriculum is represented as a directed acyclic graph in which nodes correspond to courses and edges represent prerequisite relationships. This representation enables computation of structural metrics invisible in traditional course listings: backbone courses that lie on all paths to graduation; bottleneck courses where failure concentrates; chain lengths that determine minimum time to completion; and friction indices that quantify how prerequisite structures amplify delay propagation. The graph representation transforms the curriculum from a list of requirements into a topology of constraints that shapes possible trajectories.

*Trajectory archetypes.* Rather than treating students as points on a continuous risk distribution, CAPIRE identifies discrete archetypes that capture qualitatively different patterns of progression. Using dimensionality reduction (UMAP) and clustering algorithms applied to the N1–N3 feature space, the framework identifies five recurring trajectory types: regular progression, damped progression, early friction, late exhaustion, and early exit. These archetypes differ not merely in outcome probability but in the *mechanisms* through which outcomes arise. A student exhibiting early friction faces different structural vulnerabilities than one following damped progression; interventions effective for one archetype may be irrelevant or counterproductive for another.

*Causal inference module.* CAPIRE incorporates Double Machine Learning (DML) methods to estimate causal effects under conditions of high-dimensional confounding (Chernozhukov et al., 2018). DML separates the estimation of nuisance functions (confounders) from the estimation of treatment effects, using cross-fitting to prevent overfitting and enable valid inference. The framework estimates both average treatment effects and conditional average treatment effects, revealing how interventions operate differently across archetypes, vulnerability strata, and contextual conditions. Crucially, the causal module distinguishes correlation from intervention-relevant causation: a variable may predict dropout without being a viable intervention target if it merely proxies for deeper structural factors.

### 3.2.3 Simulation layer: agent-based policy laboratory

The simulation layer implements an agent-based model (ABM) that reproduces observed trajectory patterns and enables counterfactual policy analysis. Agents represent students with heterogeneous initial conditions drawn from the empirical distribution; the environment encodes curricular structure, institutional rules, and macro-level conditions; and agent behaviour follows decision rules calibrated to observed patterns of course selection, effort allocation, and persistence.

The ABM serves two functions. First, it validates understanding: if the model reproduces observed aggregate patterns (dropout rates, time-to-degree distributions, bottleneck congestion) from micro-level rules, this provides evidence that the posited mechanisms are at least plausible. Second, it enables policy experimentation: interventions can be tested in simulation before deployment, revealing unintended consequences, adaptive responses, and interaction effects that static analysis might miss.

The simulation layer implements a factorial design crossing three intervention axes: structural changes to the curriculum (reducing prerequisite chains, adding course offerings), pedagogical changes (improving pass rates, reducing variance in course difficulty), and regulatory changes (extending regularity windows, modifying credit requirements). By simulating combinations of interventions, CAPIRE reveals which factors act as *structural amplifiers*—conditions whose modification enhances the effectiveness of other interventions—versus those that operate independently or substitutively.

### 3.3 Key empirical findings

CAPIRE's application to engineering programmes at UNT has generated findings that challenge conventional approaches to retention and illustrate the value of complexity-aware analysis.

#### 3.3.1 The curriculum as structural amplifier

Graph analysis reveals that the curriculum operates not merely as a set of requirements but as an amplifier of vulnerability. The longest prerequisite chain spans fourteen courses, establishing a structural minimum of seven semesters even under ideal conditions. Bottleneck courses—those with high betweenness centrality in the prerequisite graph—concentrate failure in ways that propagate delays across subsequent semesters. A student who fails a backbone course in the second semester may find their entire trajectory shifted, with downstream consequences that compound over years.

Critically, the structural amplification is non-linear. Small delays early in the trajectory have disproportionate effects because they interact with regularity requirements: a student who falls behind risks losing access to courses whose prerequisites they have not yet completed, which delays credit accumulation, which threatens regularity status, which further constrains course access. This positive feedback loop transforms minor setbacks into structural traps from which recovery becomes progressively more difficult.

#### 3.3.2 Archetype heterogeneity and intervention targeting

The five trajectory archetypes identified by CAPIRE exhibit distinct vulnerability profiles and respond differently to intervention. Students following regular progression face minimal structural friction and are largely insensitive to support interventions; they would succeed regardless. Students exhibiting early friction benefit substantially from targeted support in gateway courses but are insensitive to changes in upper-level pedagogy. Students in the late exhaustion archetype—who progress normally for several semesters before accumulating unsustainable burdens—require structural interventions that reduce the compounding of delay rather than individual-level support.

This heterogeneity has profound implications for policy. Uniform interventions—applied equally across archetypes—waste resources on students who do not need them and fail to reach students whose vulnerabilities lie elsewhere. Effective retention strategy requires archetype-specific targeting: matching intervention type to vulnerability mechanism rather than applying generic support across the population.

#### 3.3.3 Causal effects of friction, regularity, and macro shocks

The causal inference module estimates that curricular friction—measured as the ratio of failed prerequisites to total prerequisite load—has a substantial causal effect on dropout probability, conditional on entry characteristics and prior performance. This effect is heterogeneous: friction has minimal impact on students with strong entry profiles but large effects on students with marginal preparation. The curriculum, in other words, does not treat all students equally; it amplifies pre-existing inequalities.

Regularity status—whether a student maintains sufficient course completions to retain full enrolment privileges—operates as a threshold effect rather than a continuous risk factor. Students who lose regularity face a discrete increase in dropout hazard that cannot be explained by the underlying academic performance driving the regularity loss. The rule itself generates additional harm beyond what the rule measures.

Macro shocks interact with structural conditions in ways that produce the *dual stressor* phenomenon. Cohorts entering during periods of both high inflation and extended teachers' strikes experience dropout rates that exceed what additive models would predict. The interaction arises because both stressors reduce the margin for error—inflation by increasing opportunity costs of time, strikes by compressing instructional calendars—while the rigid curriculum provides no mechanism to absorb the resulting pressure. The system lacks resilience precisely when resilience is most needed.

#### 3.3.4 Simulation results: structural amplification of intervention effects

The agent-based model reveals that curriculum redesign operates as a structural amplifier for other interventions. Pedagogical improvements (higher pass rates) and regulatory relaxations (extended regularity windows) have modest effects when implemented in isolation against the existing curricular structure. When combined with structural redesign—shortening prerequisite chains, reducing bottleneck centrality, adding course offerings—the same interventions produce substantially larger effects.

The mechanism is straightforward: structural redesign reduces the friction that converts minor failures into cascading delays. With less friction in the system, students can absorb pedagogical variance without triggering the positive feedback loops that lead to structural exhaustion. The curriculum, when properly designed, creates slack that makes other interventions more effective.

Crucially, the simulations also reveal unintended consequences. Certain combinations of interventions—particularly those that reduce barriers without addressing underlying preparation gaps—can increase time-to-degree without reducing dropout, effectively creating a larger population of persistently struggling students. This finding illustrates why simulation matters: static analysis of intervention effects might conclude that barrier reduction is uniformly beneficial, while dynamic analysis reveals conditions under which it generates new pathologies.

### 3.4 Methodological lessons from CAPIRE development

The development of CAPIRE generated insights that extend beyond the specific domain of educational analytics. Several methodological lessons emerged from iterative engagement with data, models, and institutional stakeholders.

- *First, temporal discipline is non-negotiable.* Early versions of CAPIRE achieved impressive predictive performance by inadvertently incorporating information that would not be available at the time of intervention. Once VOT constraints were enforced, performance dropped but the resulting models were actually useful: they identified risk factors that institutions could observe and address, rather than factors that merely correlated with outcomes through unmeasured confounders.
- *Second, structure precedes prediction.* Initial attempts to predict dropout using standard machine learning approaches produced models with reasonable accuracy but limited actionability. The models could identify at-risk students but could not explain why they were at risk or what interventions might help. Graph analysis of the curriculum provided the missing explanatory structure: it revealed *where* in the system vulnerability concentrated and *how* delays propagated through prerequisite chains.
- *Third, heterogeneity is structured, not random.* Average treatment effects obscured crucial variation in how interventions operated across student populations. The archetype framework revealed that this variation was not noise but signal: different archetypes faced different vulnerability mechanisms and required different intervention strategies. Acknowledging structured heterogeneity transformed policy design from one-size-fits-all to archetype-targeted approaches.
- *Fourth, simulation reveals what static analysis conceals.* Causal estimates from observational data answer the question "what is the effect of this treatment?" but not the question "what happens when actors adapt to this policy?" The agent-based model forced explicit specification of behavioural assumptions and revealed how those assumptions shaped policy outcomes. Several interventions that appeared beneficial under static analysis generated perverse dynamics under simulation.
- *Fifth, uncertainty is irreducible but manageable.* CAPIRE does not eliminate uncertainty; it makes uncertainty explicit and navigable. The framework provides probability distributions over outcomes rather than point predictions, sensitivity analyses over modelling assumptions, and scenario comparisons that bound the range of plausible effects. This epistemic humility proved more useful to institutional decision-makers than false precision would have been.

These lessons—temporal discipline, structural priority, heterogeneity recognition, simulation necessity, uncertainty acknowledgment—emerged from the specific context of educational analytics but reflect general principles for navigating complexity. The following section extracts these principles in a form that enables their application to AI governance.

## 4. FIVE TRANSFERABLE PRINCIPLES FOR COMPLEX SYSTEMS GOVERNANCE

The methodological lessons extracted from CAPIRE's development and application can be generalised into five principles for governing complex adaptive systems. These principles are not specific to education; they address the structural challenges that any governance framework faces when applied to systems characterised by non-linearity, emergence, feedback, adaptation, and path dependence. This section articulates each principle in its general form, grounds it in CAPIRE's empirical experience, and previews its application to AI governance.

### 4.1 Principle 1: Temporal observation discipline

**Statement:** Governance frameworks must respect the temporal order in which information becomes available, distinguishing rigorously between what is known ex ante and what is learned ex post.

*Rationale*. In complex systems, the temptation to use all available information—regardless of when it becomes observable—generates systematic distortions. Models that incorporate future information into past predictions appear more accurate than they are; policies designed on such models target factors that institutions cannot actually observe at the point of intervention; and evaluations that ignore temporal sequence confuse correlation with actionable causation. Temporal discipline is not merely a methodological nicety; it is a precondition for governance that can actually be implemented.

*Grounding in CAPIRE*. The Value of Observation Time (VOT) principle emerged from CAPIRE's iterative development. Early models achieved high predictive accuracy by inadvertently incorporating N3 variables (academic trajectory) into predictions made at the N2 stage (entry moment). Once VOT constraints were enforced—ensuring that each prediction used only information available at the relevant decision point—accuracy declined but utility increased. The resulting models identified factors that admissions officers, academic advisors, and curriculum designers could actually observe and influence.

*General form*. Any governance framework operating in a complex system should: (a) explicitly map the temporal sequence in which relevant information becomes available; (b) partition variables into observation windows corresponding to distinct decision points; (c) ensure that predictive and causal models respect these partitions; and (d) evaluate interventions against information sets that were actually available when decisions were made.

*Preview for AI governance*. Risk assessment frameworks that classify AI systems based on deployment outcomes conflate ex ante evaluation with ex post observation. A complexity-aware approach would distinguish what can be assessed before deployment (architectural properties, training protocols, intended use cases) from what emerges only through deployment (actual use patterns, interaction effects, adaptive responses by users and affected populations). Governance designed for the former cannot rely on information from the latter.

### 4.2 Principle 2: Structural mapping over categorical classification

**Statement:** Governance frameworks should prioritise understanding system topology—how components connect, where constraints concentrate, how effects propagate—over assigning static categorical labels to individual elements.

*Rationale*. Categorical classification—sorting elements into risk tiers, compliance categories, or regulatory buckets—is the dominant mode of governance in both education and AI. Yet classification obscures precisely what matters most in complex systems: the relationships between elements that determine how perturbations propagate, where vulnerabilities concentrate, and which interventions might have leverage. A curriculum is not a list of courses; it is a graph of dependencies. An AI ecosystem is not a collection of systems; it is a network of interactions. Governance that addresses lists while ignoring graphs will systematically miss structural determinants of harm.

*Grounding in CAPIRE*. Graph analysis of the engineering curriculum revealed structural properties invisible in course catalogues: backbone courses that lie on all paths to graduation, bottlenecks where failure concentrates, chain lengths that establish minimum completion times, and friction indices that quantify how delays propagate. These structural properties explained variance in student outcomes that course-level attributes could not. More importantly, they identified intervention targets—reducing chain length, distributing bottleneck load, adding parallel paths—that categorical analysis would never have suggested.

*General form*. Any governance framework operating in a complex system should: (a) represent the system as a network of relationships rather than a collection of independent elements; (b) compute structural metrics that reveal concentration, connectivity, and propagation dynamics; (c) identify leverage points where intervention can reshape system topology; and (d) evaluate policies in terms of structural effects, not merely element-level compliance.

*Preview for AI governance*. The EU AI Act classifies systems into risk categories based on application domain and intended use. This categorical approach cannot capture how systems interact within decision pipelines, how risks propagate across organisational boundaries, or how compliance in one component may shift burden to others. Structural mapping would analyse AI deployment as a graph: models connected to data sources, integrated into decision processes, affecting populations who adapt in response. Governance would then target structural vulnerabilities—concentration of market power, coupling between critical systems, feedback loops between deployment and training—rather than individual system properties.

### 4.3 Principle 3: Archetype-based heterogeneity analysis

**Statement:** Governance frameworks should recognise that actors within complex systems cluster into distinct types that respond differently to identical interventions, and should design policies that account for this structured heterogeneity.

*Rationale*. Average treatment effects—the dominant metric in policy evaluation—assume that interventions operate similarly across populations. In complex adaptive systems, this assumption is routinely violated. Different actor types occupy different positions in the system structure, face different constraints, possess different adaptive capacities, and therefore respond to the same intervention in qualitatively different ways. Policies designed for the average may help no one: they may waste resources on actors who would succeed regardless, fail to reach actors whose vulnerabilities lie elsewhere, and generate perverse effects for actors whose adaptive responses differ from those anticipated.

*Grounding in CAPIRE*. The trajectory archetype framework revealed that students cluster into five distinct types—regular progression, damped progression, early friction, late exhaustion, early exit—each exhibiting different vulnerability mechanisms and intervention sensitivities. Tutoring programmes highly effective for early friction students had negligible impact on late exhaustion students, whose difficulties arose from structural accumulation rather than initial preparation gaps. Uniform retention interventions, evaluated against population averages, appeared modestly effective; archetype-specific analysis revealed that this average concealed strong positive effects for some groups and null or negative effects for others.

*General form*. Any governance framework operating in a complex system should: (a) identify recurring actor types through empirical clustering rather than imposing a priori categories; (b) estimate intervention effects separately for each archetype; (c) design differentiated policies that match intervention type to vulnerability mechanism; and (d) evaluate aggregate outcomes as weighted combinations of archetype-specific effects, not as homogeneous population responses.

*Preview for AI governance*. The AI ecosystem contains heterogeneous actor types—large technology firms, startups, academic researchers, open-source communities, public sector deployers—that differ systematically in compliance capacity, market position, adaptive resources, and vulnerability to regulatory pressure. Uniform regulations may entrench incumbents (who can absorb compliance costs), exclude new entrants (who cannot), marginalise non-commercial development (which lacks resources for certification), and displace activity to less regulated contexts. Archetype-aware governance would analyse how different actor types respond to proposed regulations and design differentiated obligations that achieve policy objectives without generating structural distortions.

### 4.4 Principle 4: Causal mechanism identification

**Statement:** Governance frameworks should distinguish correlation from intervention-relevant causation, identifying the mechanisms through which effects arise rather than merely the statistical associations that predict outcomes.

*Rationale*. Prediction and intervention require different kinds of knowledge. A variable may predict an outcome without being a viable intervention target: it may proxy for unmeasured confounders, reflect downstream effects rather than upstream causes, or operate through mechanisms that policy cannot influence. Conversely, a variable may have modest predictive power yet substantial causal leverage if it lies upstream of amplifying feedback loops. Governance based on predictive associations will systematically misallocate intervention resources, targeting symptoms while leaving causes unaddressed or generating iatrogenic effects by intervening on correlates rather than causes.

*Grounding in CAPIRE*. The Double Machine Learning module estimated causal effects of curricular friction, regularity status, and macro shocks on dropout probability. Several strong predictors—such as first-semester GPA—proved to have limited causal leverage: they primarily reflected pre-entry preparation differences that the university could not modify. Conversely, regularity status—a policy-determined threshold—had causal effects beyond what underlying performance would predict, indicating that the rule itself generated additional harm amenable to policy intervention. Without causal analysis, intervention would have targeted the wrong factors.

*General form*. Any governance framework operating in a complex system should: (a) employ causal inference methods that distinguish correlation from causation under conditions of confounding; (b) estimate both average and heterogeneous treatment effects to reveal differential mechanism operation; (c) articulate explicit causal models that specify the pathways through which interventions are expected to operate; and (d) update causal understanding based on intervention outcomes, distinguishing model failure from implementation failure.

*Preview for AI governance*. Much AI governance discourse conflates correlation (AI deployment is associated with certain harms) with causation (AI caused those harms). Discriminatory outcomes may arise from AI systems, from the data environments in which they operate, from the institutional contexts of deployment, or from pre-existing social conditions that AI merely reveals. Effective governance requires disentangling these causal pathways: interventions targeting model properties will fail if harm arises primarily from deployment context; interventions targeting deployment will fail if harm is baked into training data; interventions targeting data will fail if harm reflects structural inequalities that no technical fix can address. Causal mechanism identification is prerequisite to intervention design.

### 4.5 Principle 5: Simulation-based policy design

**Statement:** Governance frameworks should test proposed interventions against simulated adaptive responses before deployment, exploring unintended consequences, interaction effects, and emergent dynamics that static analysis cannot reveal.

*Rationale*. In complex adaptive systems, actors respond strategically to governance interventions. Regulations become part of the environment to which firms, users, and institutions adapt. Static analysis—estimating intervention effects while holding behaviour constant—systematically underestimates adaptive responses, fails to anticipate emergent configurations, and misses interaction effects between simultaneous interventions. Simulation forces explicit specification of behavioural assumptions, reveals how those assumptions shape outcomes, and enables exploration of policy combinations before committing to implementation.

*Grounding in CAPIRE*. The agent-based model revealed that curriculum redesign operates as a structural amplifier for other interventions. Pedagogical improvements and regulatory relaxations had modest effects in isolation but substantially larger effects when combined with structural changes that reduced friction accumulation. The simulation also revealed unintended consequences: certain intervention combinations increased time-to-degree without reducing dropout, creating a larger population of struggling persisters. These dynamics were invisible to static analysis; they emerged only when adaptive behaviour was explicitly modelled.

*General form*. Any governance framework operating in a complex system should: (a) construct simulation models that represent agent heterogeneity, decision rules, and environmental constraints; (b) calibrate simulations against observed system behaviour to establish baseline validity; (c) test proposed interventions in simulation before deployment, exploring parameter sensitivity and scenario variation; (d) use simulation to identify structural amplifiers—interventions whose presence enhances the effectiveness of others; and (e) treat simulation results as hypotheses to be tested, not predictions to be trusted, updating models as implementation evidence accumulates.

*Preview for AI governance*. AI governance frameworks are currently designed without systematic simulation of adaptive responses. How will firms restructure operations to minimise regulatory exposure? How will compliance requirements affect market structure? How will classification thresholds shape development trajectories? These questions are answerable—at least provisionally—through agent-based models that represent firm behaviour, market dynamics, and regulatory interaction. Simulation would not provide certainty, but it would reveal possibilities that purely conceptual analysis cannot anticipate, enabling governance design that accounts for adaptation rather than assuming compliance.

### 4.6 Integration: from principles to framework

The five principles are not independent prescriptions but integrated components of a coherent approach to complex systems governance. Temporal discipline ensures that analysis uses information actually available at decision points. Structural mapping reveals the topology within which interventions operate. Archetype analysis identifies differentiated responses to policy. Causal identification distinguishes actionable levers from mere correlates. Simulation tests interventions against adaptive dynamics before commitment.

Together, these principles constitute a methodological stance: governance as intervention design under uncertainty, informed by structural understanding, tested through simulation, and revised through learning. The following section applies this stance to AI governance, proposing Complex Systems AI Governance (CSAIG) as an integrated framework that operationalises the five principles for regulatory design.

## 5. APPLICATION: REDESIGNING AI GOVERNANCE THROUGH A COMPLEXITY LENS

The five principles extracted from CAPIRE provide a template for complexity-aware governance. This section applies that template to artificial intelligence, first critiquing existing frameworks against the principles, then proposing Complex Systems AI Governance (CSAIG) as an alternative architecture, and finally illustrating CSAIG's operation through a concrete scenario.

### 5.1 Critique of existing frameworks

Three regulatory frameworks currently dominate AI governance discourse: the European Union's Artificial Intelligence Act (EU AI Act), the United States' AI Risk Management Framework (NIST AI RMF), and the OECD's Recommendation on Artificial Intelligence. Each represents a serious attempt to address AI risks; each falls short when evaluated against complexity principles.

#### 5.1.1 Temporal discipline violations

All three frameworks conflate ex ante assessment with ex post outcomes. The EU AI Act classifies systems by intended use and application domain, implicitly assuming that pre-deployment characteristics predict post-deployment harms. Yet the most consequential harms often emerge through deployment dynamics that cannot be anticipated from system properties alone: user adaptation, interaction with other systems, feedback between model outputs and training data, and strategic responses by affected populations (Selbst et al., 2019).

The NIST framework acknowledges uncertainty but provides no systematic method for distinguishing observable from latent risks at different lifecycle stages. Risk assessments that incorporate deployment information into pre-deployment decisions generate the same temporal leakage that undermined early CAPIRE models: inflated confidence in assessments that cannot actually be made when decisions must be taken.

#### 5.1.2 Categorical classification without structural mapping

The EU AI Act's tiered risk classification exemplifies categorical thinking. Systems are sorted into buckets—unacceptable, high, limited, minimal risk—based on application domain and intended use. This classification cannot capture structural properties of the AI ecosystem: how systems connect within decision pipelines, where market concentration creates systemic vulnerabilities, how data flows link nominally independent applications, or which architectural choices create path dependencies that constrain future options.

The analogy to CAPIRE is direct. Classifying courses by difficulty level would miss the structural properties—prerequisite chains, bottleneck positions, path dependencies—that actually determine student outcomes. Similarly, classifying AI systems by risk tier misses the relational properties that determine ecosystem dynamics. A 'limited risk' system that feeds into multiple high-risk decision pipelines may have greater systemic impact than a nominally ‘high-risk’ system operating in isolation.

#### 5.1.3 Homogeneous treatment of heterogeneous actors

Existing frameworks apply uniform obligations across actor types, despite vast differences in compliance capacity, market position, and adaptive resources. The EU AI Act imposes the same conformity assessment requirements on multinational technology firms and university research groups, on well-resourced startups and open-source communities, on commercial deployers and public interest applications.

CAPIRE’s archetype analysis suggests this uniformity will produce structural distortions. Actors with high compliance capacity (large firms) will absorb requirements and potentially benefit from barriers to entry. Actors with limited resources (startups, academics, civil society) will be excluded from regulated domains or pushed toward unregulated alternatives. The distribution of AI development will shift in ways that may concentrate rather than distribute the risks regulation seeks to address.

#### 5.1.4 Correlational rather than causal reasoning

Current frameworks identify risk factors through association rather than causal analysis. Certain application domains are deemed high-risk because past deployments in those domains have generated harm. But this correlational reasoning cannot distinguish whether harm arose from AI system properties, deployment contexts, institutional failures, or pre-existing social conditions that AI merely made visible.

The distinction matters for intervention design. If discriminatory hiring outcomes arise primarily from biased training data, then model-level regulation will be insufficient. If they arise from institutional practices that AI automates, then data-level interventions will fail. If they reflect structural labour market inequalities, then technical governance of any kind may be beside the point. Without causal mechanism identification, regulatory effort targets the most visible correlate rather than the most tractable cause.

#### 5.1.5 Absence of simulation and adaptive anticipation

None of the major frameworks incorporates systematic simulation of regulatory effects. The EU AI Act was developed through consultation, impact assessment, and legal analysis—but not through computational exploration of how actors might adapt to proposed requirements. How will firms restructure to minimise classification as high-risk? How will compliance markets evolve? How will regulatory arbitrage reshape the geographic distribution of AI development?

These questions are not merely speculative; they are answerable through simulation methods analogous to those CAPIRE employed for educational policy. The absence of such simulation means that AI governance is implemented as an uncontrolled experiment: policymakers propose interventions, observe outcomes, and adjust—but without the systematic exploration of possibility space that would enable anticipation of adaptive dynamics.

### 5.2 Complex Systems AI Governance (CSAIG): an alternative architecture

CSAIG operationalises the five complexity principles for AI governance. It is not a complete regulatory framework—specifying detailed obligations, enforcement mechanisms, and institutional arrangements would exceed the scope of this paper and require domain expertise beyond what educational analytics can provide. Rather, CSAIG is an *architectural template*: a specification of the analytical components that complexity-aware AI governance must include, regardless of the specific policy content those components support.

Table 2 summarises the mapping from complexity principles to CSAIG components.

**Table 2.** Mapping from complexity principles to CSAIG components

| Principle | CSAIG component | Function |
|---|---|---|
| Temporal discipline | Lifecycle observation mapping | Specifies what is observable at each governance decision point |
| Structural mapping | Ecosystem topology analysis | Maps connections, dependencies, and propagation pathways |
| Archetype analysis | Actor type differentiation | Identifies distinct actor types and their differential responses |
| Causal identification | Mechanism specification module | Distinguishes causal pathways from correlational associations |
| Simulation | Policy simulation laboratory | Tests interventions against adaptive responses before deployment |
| (Integration) | Adaptive governance protocol | Institutionalises monitoring, learning, and revision cycles |

### 5.2.1 Lifecycle observation mapping

CSAIG's first component partitions the AI lifecycle into observation windows analogous to CAPIRE's N1–N4 levels:

*Design stage:* Information available before deployment—architectural choices, training data characteristics, intended use specifications, documented capabilities and limitations.

*Deployment stage:* Information generated at the point of release—deployment context, user population, integration with existing systems, initial monitoring arrangements.

*Operation stage:* Information accumulated during use—actual use patterns, performance metrics, incident reports, user feedback, model drift indicators.

*Ecosystem stage:* Information about system-level conditions—market structure, competitive dynamics, regulatory responses in other jurisdictions, technological developments that alter the landscape.

Governance obligations are calibrated to observation windows: design-stage assessments use only design-stage information; deployment decisions incorporate deployment-stage information; operational oversight uses operational data; and ecosystem-level governance addresses ecosystem-level dynamics. This calibration prevents the temporal leakage that undermines assessment validity.

#### 5.2.2 Ecosystem topology analysis

CSAIG's second component maps the AI ecosystem as a network of relationships rather than a collection of independent systems. The mapping includes:

*Decision pipeline graphs:* How AI systems connect within organisational decision processes—which systems feed into which decisions, where human oversight is inserted, how outputs from one system become inputs to another.

*Data flow networks:* How data moves through the ecosystem—training data provenance, inference data collection, feedback loops between deployment outputs and model updates.

*Market structure maps:* Concentration, vertical integration, and dependency relationships that create systemic vulnerabilities—cloud infrastructure dominance, model API dependencies, data broker networks.

*Regulatory coupling analysis:* How governance in one domain interacts with governance in others—data protection requirements affecting AI training, sector-specific regulations constraining deployment options, liability regimes shaping development incentives.

From these mappings, CSAIG computes structural metrics analogous to those CAPIRE derives from curriculum graphs: concentration indices that identify systemically important actors, coupling measures that reveal propagation pathways, bottleneck analyses that locate vulnerabilities, and leverage point identification that suggests where intervention might have disproportionate effect.

#### 5.2.3 Actor type differentiation

CSAIG's third component identifies distinct actor types through empirical analysis rather than a priori categorisation. Initial typologies might distinguish:

*Large integrated firms:* High compliance capacity, market power, resources for regulatory engagement, incentives toward regulatory capture.

*Venture-backed startups:* Moderate resources, time pressure, sensitivity to compliance costs, potential for acquisition by larger firms.

*Academic researchers:* Limited deployment scale, publication incentives, institutional review processes, resource constraints.

*Open-source communities:* Distributed development, unclear liability, minimal formal compliance capacity, ideological motivations.

*Public sector deployers:* Accountability requirements, procurement constraints, democratic legitimacy concerns, limited technical expertise.

For each actor type, CSAIG estimates differential responses to proposed regulations: compliance costs, adaptive strategies, exit options, and systemic effects. This analysis enables differentiated policy design that achieves objectives without generating structural distortions—for instance, tiered obligations based on deployment scale, safe harbours for research applications, or compliance support for resource-constrained public interest uses.

### 5.2.4 Mechanism specification module

CSAIG's fourth component requires explicit causal models for the harms that governance seeks to address. For each harm category—discrimination, privacy violation, safety failure, market distortion—the module specifies:

*Causal pathways:* Through which mechanisms does the harm arise? Training data bias? Deployment context? Institutional practice? Structural inequality? Multiple interacting causes?

*Intervention points:* Where in the causal chain can governance act? Which interventions target tractable causes versus immutable correlates?

*Effect heterogeneity:* How do causal effects vary across contexts, populations, and deployment conditions? Where are interventions most and least effective?

*Evidence requirements:* What evidence would distinguish between competing causal accounts? What data would update beliefs about mechanism operation?

This explicit causal modelling prevents governance from targeting visible correlates while leaving causes unaddressed. It also enables rational prioritisation: interventions are evaluated not by political salience but by causal leverage.

### 5.2.5 Policy simulation laboratory

CSAIG's fifth component implements agent-based simulation of regulatory effects. The simulation represents:

*Agent populations:* Heterogeneous actors with type-specific characteristics, decision rules, and adaptive capacities, calibrated to empirical distributions.

*Environmental structure:* Market conditions, technological constraints, existing regulatory requirements, and information flows that shape agent behaviour.

*Regulatory interventions:* Proposed policies implemented as modifications to environmental constraints, cost structures, or compliance requirements.

*Outcome metrics:* Measures of policy objectives (harm reduction, innovation effects, equity impacts) and unintended consequences (market concentration, regulatory arbitrage, compliance theatre).

The laboratory enables systematic exploration of policy combinations, revealing structural amplifiers (interventions whose presence enhances others) and structural conflicts (interventions that undermine each other). It also identifies conditions under which policies generate perverse effects—increasing the harms they seek to reduce or creating new pathologies not present in the unregulated baseline.

#### 5.2.6 Adaptive governance protocol

CSAIG's sixth component institutionalises ongoing learning. Recognising that governance in complex systems cannot be designed once and implemented forever, the protocol establishes:

*Monitoring systems:* Continuous collection of data on policy implementation, actor responses, and outcome trajectories.

*Evaluation frameworks:* Methods for assessing whether observed outcomes match simulation predictions and causal models, with explicit criteria for updating beliefs.

*Revision mechanisms:* Institutional procedures for modifying policies in response to evidence, balancing stability with adaptability.

*Sunset and review clauses:* Automatic reassessment triggers that prevent obsolete regulations from persisting through inertia.

The adaptive protocol treats governance as an ongoing experiment rather than a settled solution. This stance is not a concession to regulatory weakness but a recognition of epistemic reality: in complex systems, certainty is unattainable and the most dangerous governance is that which pretends otherwise.

### 5.3 Illustrative scenario: regulatory fragmentation revisited

To illustrate CSAIG's operation, consider the regulatory fragmentation scenario introduced earlier (Paz, 2025b): a jurisdiction implements risk-based AI regulation requiring high-risk systems to meet stringent conformity assessment requirements. Firms respond by restructuring decision pipelines, distributing functionality across multiple components such that no single component triggers classification thresholds. Formal compliance increases while harm persists or migrates.

Under current frameworks, this outcome would emerge as a surprise—discovered only after implementation, when the gap between compliance metrics and harm indicators becomes impossible to ignore. CSAIG would approach the scenario differently at each stage.

*Lifecycle observation mapping* would distinguish what can be assessed before deployment (component properties, integration architecture) from what emerges through deployment (actual decision pathways, human oversight effectiveness). Risk assessment would not conflate design-stage characteristics with operational outcomes.

*Ecosystem topology analysis* would map decision pipelines rather than classifying individual components. The analysis would reveal that regulatory exposure depends on pipeline structure, not component properties—creating incentives for fragmentation that categorical classification cannot address.

*Actor type differentiation* would identify which firms possess the compliance capacity and architectural flexibility to restructure pipelines, and which would bear compliance costs without adaptive options. This analysis would predict distributional effects invisible to uniform regulatory design.

*Mechanism specification* would articulate the causal pathways through which harm arises—distinguishing component-level failures from integration failures from contextual factors. This would reveal whether component-focused regulation can address integration-level harms.

*Policy simulation* would model firm responses to classification thresholds, revealing fragmentation incentives before implementation. Simulation would explore alternative designs—pipeline-level assessment, integration audits, outcome-based rather than component-based obligations—and identify which approaches are robust to adaptive responses.

*Adaptive governance* would establish monitoring for fragmentation indicators and revision triggers when compliance metrics diverge from harm indicators. The framework would treat the initial regulation as a hypothesis to be tested, not a solution to be defended.

CSAIG does not guarantee that fragmentation would be prevented—complex systems offer no such guarantees. But it would ensure that fragmentation is *anticipated*, that alternatives are *explored*, and that governance is *prepared to adapt* when predictions prove wrong. This is what complexity-aware governance can offer: not certainty but informed navigation of uncertainty.

### 5.4 Limitations of the transfer

The transfer from CAPIRE to CSAIG rests on structural isomorphism, but isomorphism is not identity. Several limitations constrain the extent to which educational analytics lessons apply to AI governance.

*Scale and observability*. Educational systems generate dense longitudinal data on individual trajectories; AI ecosystems generate sparser, more fragmented data with greater commercial sensitivity. The data infrastructure that enabled CAPIRE's empirical grounding does not yet exist for AI governance. CSAIG's components are therefore specified abstractly, awaiting the development of monitoring systems that could populate them with empirical content.

*Temporal dynamics*. Student trajectories unfold over years; AI system deployment effects may emerge over weeks or persist for decades. The temporal scales of adaptation differ: curricular responses to student behaviour occur semesterly; firm responses to regulation may occur continuously. CSAIG must accommodate faster feedback loops and longer path dependencies than CAPIRE was designed to address.

*Stakes and reversibility*. Failed retention interventions harm students; failed AI governance may harm populations at scale. Educational policy errors are often reversible; regulatory lock-in may persist across technological generations. The asymmetry in stakes argues for greater caution in AI governance—but also for greater investment in simulation that could reduce reliance on learning from deployed failure.

*Political economy.* Universities have limited resources but also limited opposition to retention improvement. AI governance faces well-resourced actors with strong incentives for regulatory capture, jurisdictional arbitrage, and political influence. The adaptive responses that CSAIG models may include not merely operational restructuring but strategic engagement with the governance process itself.

These limitations do not invalidate the transfer; they bound its applicability. CSAIG offers an architectural template grounded in complexity principles—but the specific implementation must account for features of the AI domain that educational analytics did not confront. The next section discusses implications for policy, institutions, and research given these constraints.

## 6. DISCUSSION

The preceding sections have argued that complexity constitutes a shared ontology linking educational systems and AI ecosystems, extracted five transferable principles from CAPIRE's empirical development, and proposed CSAIG as an architectural template for complexity-aware AI governance. This section reflects on the implications of this argument for theory, practice, and future research, while acknowledging the constraints within which any such contribution operates.

### 6.1 Theoretical contributions

The paper's primary theoretical contribution is demonstrating the feasibility and value of cross-domain transfer in complex systems governance. While the application of complexity theory to both education (Jacobson & Wilensky, 2006; Mason, 2008) and AI governance (Cath, 2018; Yeung, 2020) has been proposed independently, the systematic derivation of transferable principles from one domain to the other represents a novel methodological move.

This move rests on a claim that deserves explicit articulation: that complex adaptive systems, despite their domain-specific content, share structural properties that make governance challenges *formally similar* across domains. The five properties—non-linearity, emergence, feedback, adaptation, path dependence—are not merely analogies but structural isomorphisms that determine how systems respond to intervention. If this claim is correct, then the methodological innovations developed in any complex systems domain are, in principle, portable to others.

The claim has limits. Structural isomorphism does not imply identical dynamics; the specific constants, timescales, and actor characteristics differ across domains. Transfer requires translation, not mere application. But the theoretical contribution remains: frameworks for navigating complexity need not be invented de novo for each domain. Learning can cumulate across fields.

A secondary theoretical contribution is the articulation of the five principles themselves. While each principle draws on established methodological traditions—temporal logic in causal inference (Imbens & Rubin, 2015), network analysis in organisational studies (Borgatti et al., 2009), heterogeneous treatment effects in econometrics (Athey & Imbens, 2016), agent-based modelling in policy analysis (Gilbert, 2008)—their integration into a coherent framework for complex systems governance is, to our knowledge, novel. The principles are not merely a list but a system: temporal discipline enables structural mapping; structural mapping reveals heterogeneity; heterogeneity analysis informs causal identification; causal identification grounds simulation; simulation enables adaptive governance. Each principle supports and constrains the others.

### 6.2 Implications for regulatory practice

If the argument presented here is sound, what would change in regulatory practice? Several implications follow.

*First, classification would be supplemented by mapping.* Risk-based classification need not be abandoned, but it should be complemented by structural analysis of the regulated ecosystem. Regulators would invest in understanding how systems connect, where dependencies concentrate, and how effects propagate—not merely in sorting systems into risk tiers. This implies new regulatory capacities: network analysis expertise, data infrastructure for ecosystem mapping, and institutional arrangements for updating maps as the ecosystem evolves.

*Second, impact assessment would incorporate simulation.* Before implementing significant regulatory changes, agencies would explore adaptive responses through computational models. These simulations would not provide certainty—their value lies precisely in revealing uncertainties that conceptual analysis might miss. Regulatory proposals would be accompanied by simulation results showing expected outcomes under different behavioural assumptions, sensitivity analyses identifying conditions under which effects reverse, and explicit acknowledgment of what the simulations cannot capture.

*Third, obligations would be differentiated by actor type.* Rather than uniform requirements applied regardless of compliance capacity, market position, or adaptive resources, regulations would incorporate type-specific provisions. This differentiation need not imply weaker oversight for powerful actors; it might mean stronger structural requirements for concentrated markets, safe harbours for research applications, and compliance support for public interest uses. The goal is achieving policy objectives without generating structural distortions that concentrate rather than distribute risk.

*Fourth, monitoring would focus on system-level outcomes rather than component-level compliance.* Regulatory success would be evaluated against harm indicators, market structure metrics, and innovation effects—not merely against conformity assessment completion rates. When compliance metrics diverge from outcome metrics, this would trigger investigation and potential revision rather than continued reliance on procedural proxies.

*Fifth, governance would be explicitly adaptive.* Regulations would include sunset clauses, review triggers, and revision mechanisms that make policy adjustment routine rather than exceptional. Regulators would communicate uncertainty openly, treating initial rules as informed hypotheses rather than settled solutions. This epistemic humility might initially seem to weaken regulatory authority; in practice, it would enhance credibility by aligning regulatory claims with regulatory capacity.

### 6.3 Implications for institutions

Implementing complexity-aware governance requires institutional capacities that most regulatory agencies currently lack. The implications extend beyond technical expertise to organisational culture and inter-institutional relationships.

*Capacity building.* Regulatory agencies would need expertise in network analysis, causal inference, and agent-based modelling—fields rarely represented in legal and policy workforces. This implies either substantial internal hiring or systematic collaboration with academic and research institutions. Neither option is straightforward: internal hiring competes with private sector salaries; external collaboration raises questions of independence and capture. But the alternative—governing complex systems without the tools to understand them—is worse.

*Cultural transformation.* Complexity-aware governance requires comfort with uncertainty that many regulatory cultures resist. Agencies accustomed to projecting confidence and control must learn to communicate probabilistically, acknowledge limitations, and revise positions in response to evidence. This transformation faces internal resistance from staff trained in deterministic legal reasoning and external pressure from political principals who prefer simple narratives. Yet the alternative—false certainty that erodes credibility when predictions fail—is unsustainable.

*Inter-institutional coordination.* AI ecosystems span jurisdictional and sectoral boundaries. No single regulator possesses the authority, data, or expertise to implement CSAIG alone. Effective governance requires coordination mechanisms that share structural maps, align actor typologies, reconcile causal models, and integrate simulation results across agencies and jurisdictions. Existing coordination mechanisms—mutual recognition agreements, information sharing protocols, harmonisation initiatives—would need substantial enhancement to support complexity-aware governance.

*Democratic accountability.* Complexity-aware governance involves technical methods that may be opaque to democratic oversight. Simulations rest on assumptions that embed values; archetype classifications shape who bears regulatory burden; causal models privilege certain intervention pathways over others. These choices are not merely technical; they are political. Institutional designs must ensure that complexity-aware governance remains accountable to democratic processes, with transparent methodologies, contestable assumptions, and accessible explanations of how conclusions were reached.

### 6.4 Implications for research

The argument presented here opens several research agendas that extend beyond what this paper could address.

*Empirical validation of CSAIG components*. CSAIG is currently an architectural template, not an implemented system. Research is needed to develop specific methods for AI ecosystem topology analysis, to construct and validate actor typologies through empirical clustering, to build causal models of AI harm pathways, and to calibrate agent-based simulations against observed regulatory responses. Each component requires methodological development and empirical grounding.

*Cross-domain transfer methodology*. This paper demonstrates one instance of cross-domain transfer—from education to AI governance. Research should examine whether similar transfers are possible from other complex systems domains: healthcare governance, financial regulation, environmental management, urban planning. What are the conditions under which transfer succeeds? What domain-specific adaptations are required? How can transfer methodology be systematised?

*Simulation validity in policy contexts*. Agent-based models are powerful but not infallible. Research is needed on how to assess simulation validity when ground truth is unavailable, how to communicate simulation uncertainty to policymakers, and how to update models as implementation evidence accumulates. The epistemology of simulation-based governance remains underdeveloped.

*Institutional design for adaptive governance*. How should regulatory institutions be structured to enable continuous learning and revision? What legal frameworks support adaptive governance while maintaining rule-of-law values? How can revision mechanisms avoid capture by regulated actors? These questions lie at the intersection of administrative law, institutional economics, and organisational theory.

*Political economy of complexity-aware regulation*. Complexity-aware governance may threaten actors who benefit from regulatory simplicity—whether through compliance capacity advantages or through the predictability that enables regulatory arbitrage. Research should examine the political conditions under which complexity-aware approaches can be adopted, the coalitions that might support or oppose them, and the strategies for overcoming resistance.

### 6.5 Limitations of this study

Several limitations constrain the conclusions that can be drawn from this analysis.

*Single-case empirical foundation*. CAPIRE was developed and validated in a single institutional context—engineering programmes at one Argentine university. While the methodological principles are articulated in general terms, their empirical grounding is narrow. Replication across educational contexts and validation in AI governance contexts would strengthen confidence in transferability.

*Conceptual rather than implemented transfer*. CSAIG remains an architectural proposal, not an implemented system. The paper demonstrates that transfer is conceptually coherent but not that it is practically feasible. Implementation would reveal challenges that conceptual analysis cannot anticipate.

*Author positioning*. The author developed both CAPIRE and the argument for its transfer to AI governance. This positioning creates potential for confirmation bias—overestimating CAPIRE's achievements and the relevance of its lessons. Independent evaluation by researchers without stake in either framework would provide valuable correction.

*Disciplinary constraints*. The paper approaches AI governance from the perspective of complex systems methodology and educational analytics. It does not engage deeply with legal scholarship on regulatory design, political science perspectives on governance, or economic analysis of regulatory effects. A fuller treatment would integrate these disciplinary perspectives.

These limitations do not invalidate the argument but bound its scope. The paper offers a methodological contribution—demonstrating that complexity principles developed in education can inform AI governance—not a complete governance framework. Further development requires collaboration across disciplines and validation through implementation.

## 7. CONCLUSION

### 7.1 Summary of the argument

This paper began with an observation: two domains that rarely appear in the same conversation—student retention in higher education and artificial intelligence governance—share a structural pathology. Both are governed by frameworks designed for linear systems yet operate as complex adaptive systems where linearity assumptions systematically fail. Risk-based approaches dominate both domains, yet routinely underperform because they assume stable causal pathways, predictable actor responses, and controllable system boundaries.

From this observation, we developed an argument in four stages. First, we established that complexity constitutes a shared ontology linking educational systems and AI ecosystems. Both exhibit non-linearity, emergence, feedback loops, strategic adaptation, and path dependence—the defining properties of complex adaptive systems. This structural isomorphism justifies treating methodological principles as transferable rather than domain-specific.

Second, we presented CAPIRE, a framework developed and empirically validated for analysing student trajectories in engineering education. CAPIRE treats dropout as an emergent property of curricular structure, institutional rules, and external shocks. Its architecture—multi-level data organisation, graph-theoretic curriculum analysis, trajectory archetype identification, causal inference through Double Machine Learning, and agent-based policy simulation—emerged from iterative engagement with real institutional data and generated findings that challenged conventional retention approaches.

Third, we extracted five transferable principles from CAPIRE's development: temporal observation discipline, structural mapping over categorical classification, archetype-based heterogeneity analysis, causal mechanism identification, and simulation-based policy design. Each principle addresses a structural challenge that any governance framework faces when applied to complex adaptive systems.

Fourth, we applied these principles to AI governance, critiquing existing frameworks against complexity criteria and proposing Complex Systems AI Governance (CSAIG) as an architectural template. CSAIG operationalises the five principles through six integrated components: lifecycle observation mapping, ecosystem topology analysis, actor type differentiation, mechanism specification, policy simulation laboratory, and adaptive governance protocol.

### 7.2 The core claim

The paper's core claim is that the central question of governance must shift. In education, the question should not be 'which students are at risk?' but 'how do curricular structures and institutional rules generate vulnerability, and how do interventions reshape those dynamics?' In AI governance, the question should not be 'how risky is this system?' but 'how does this regulatory intervention alter the behaviour of the sociotechnical system, and what new configurations might it generate?'

This shift is not merely semantic. It implies different analytical methods (structural mapping rather than categorical classification), different evaluation criteria (system-level outcomes rather than component-level compliance), different institutional capacities (simulation and adaptation rather than static rule application), and different epistemic stances (acknowledged uncertainty rather than projected control).

The shift also implies different relationships between governors and governed. Linear governance positions regulators as external controllers applying constraints to a passive system. Complexity-aware governance positions regulators as participants in an evolving system, intervening strategically but unable to dictate outcomes, learning from responses, and adapting as the system co-evolves with its governance regime. This positioning is more humble but also more honest: it aligns regulatory claims with regulatory capacity.

### 7.3 What complexity-aware governance can and cannot offer

Complexity-aware governance does not promise certainty, control, or elimination of harm. Complex adaptive systems are, by their nature, partially unpredictable. Emergence means that novel configurations will arise that no analysis anticipated. Adaptation means that actors will find ways around constraints that governance imposes. Path dependence means that some trajectories, once established, cannot be reversed regardless of subsequent intervention.

What complexity-aware governance offers instead is *informed navigation of uncertainty*. Structural mapping reveals where vulnerabilities concentrate and where intervention might have leverage. Archetype analysis identifies differentiated responses that uniform approaches would miss. Causal identification distinguishes actionable factors from mere correlates. Simulation explores adaptive responses before commitment to implementation. Adaptive protocols enable learning and revision as evidence accumulates.

This offering is less satisfying than the promise of control that linear frameworks implicitly make. But the promise of control is false. Linear frameworks do not actually deliver certainty; they merely project it, generating confidence that erodes when predictions fail. Complexity-aware governance trades false certainty for genuine capability: the capability to anticipate more possibilities, to prepare for more contingencies, and to adapt more effectively when surprises occur.

### 7.4 The value of cross-domain learning

Beyond its specific contributions to AI governance, this paper illustrates a broader methodological point: that learning can transfer across domains when those domains share structural properties. Educational analytics and AI governance are not obviously related fields. Their literatures rarely cite each other; their practitioners rarely interact; their institutional homes are entirely separate. Yet the structural isomorphism between educational systems and AI ecosystems means that methodological innovations in one domain can accelerate development in the other.

This point generalises. Healthcare systems, financial markets, urban environments, ecological networks—all exhibit complex adaptive system properties. Governance innovations developed in any of these domains may, with appropriate translation, inform governance in the others. The costs of complexity-aware governance—capacity building, methodological development, institutional transformation—need not be paid independently by each domain. Cross-domain learning enables sharing of these costs and acceleration of collective capability.

Realising this potential requires institutional arrangements that facilitate cross-domain communication: interdisciplinary research programmes, boundary-spanning professional communities, and publication venues that welcome methodological transfers. It also requires intellectual humility: willingness to learn from distant fields, to translate rather than merely apply, and to acknowledge that no single domain has solved the challenge of governing complexity.

### 7.5 Final reflection

The challenge of governing artificial intelligence is often framed as unprecedented—a novel problem requiring novel solutions. There is truth in this framing: AI systems possess capabilities, scale, and speed that distinguish them from prior technologies. But there is also distortion. The *structural* challenge of governing AI—intervening in a complex adaptive system characterised by non-linearity, emergence, feedback, adaptation, and path dependence—is not novel at all. Humanity has faced this challenge in domains ranging from ecosystems to economies, from public health to urban development. We have accumulated, across these domains, substantial experience with what works and what fails when linear governance confronts complex systems.

This paper argues that we should draw on that experience. CAPIRE was developed in the modest context of engineering education at a single Argentine university, addressing the unglamorous problem of student dropout. Yet the methodological innovations it required—temporal discipline, structural mapping, archetype analysis, causal identification, simulation-based design—address challenges that AI governance faces in amplified form. The transfer is not perfect; translation is required; limitations must be acknowledged. But the alternative—reinventing complexity-aware governance from scratch for each new domain—wastes accumulated knowledge and delays the development of adequate governance capacity.

Artificial intelligence will reshape societies in ways we cannot fully anticipate. Governing this transformation requires frameworks adequate to its complexity—frameworks that acknowledge uncertainty, map structure, differentiate responses, identify mechanisms, test interventions through simulation, and adapt as evidence accumulates. Such frameworks exist, developed through decades of engagement with complex systems across many domains. The task is not to invent them but to translate, integrate, and apply them. This paper offers one contribution to that task. Many more will be needed.